\documentclass[a4paper,11pt,dvips]{article}
\usepackage{graphicx}
\usepackage{amsmath}
\usepackage{amssymb}
\usepackage{latexsym}
\usepackage[colorlinks]{hyperref}
\usepackage{color}
\usepackage{cite}

\newcommand{\bra}{\begin{array}}
\newcommand{\era}{\end{array}}
\newcommand{\beq}{\begin{equation}}
\newcommand{\eeq}{\end{equation}}
\newcommand{\beqar}{\begin{eqnarray}}
\newcommand{\eeqar}{\end{eqnarray}}

\def\BC{\bb C}
\def\_\BC{\bbi C}

\def\( {\left(}
   \def\) {\right)}
\def\[ {\left[}
\def\] {\right]}
\def\no2 {{\textstyle{n\over 2}}}

\newcommand{\lb}{\label}

\begin{document}
\begin{titlepage}
\setcounter{page}{1}
\renewcommand{\thefootnote}{\fnsymbol{footnote}}

\begin{flushright}
\end{flushright}

\vspace{5mm}
\begin{center}

{\Large \bf {Energy Levels of an Ideal Quantum Ring \\ in
AA-Stacked Bilayer Graphene}}

\vspace{5mm}

{\bf Youness Zahidi$^{a,b}$, Abdelhadi Belouad$^{b}$} and {\bf Ahmed
Jellal\footnote{\sf ajellal@ictp.it --
jellal.a@ucd.ac.ma}}$^{b,c}$

\vspace{5mm}
{$^{a}$\em  MATIC, FPK, Hassan 1 University, Khouribga, Morocco}\\
{$^{b}$\em Theoretical Physics Group,  
Faculty of Sciences, Choua\"ib Doukkali University},\\
{\em PO Box 20, 24000 El Jadida, Morocco} \\
{$^c$\em Saudi Center
for Theoretical Physics, Dhahran, Saudi Arabia}


\vspace{3cm}

\begin{abstract}

We theoretically analyze  the energy spectrum of a quantum ring in
AA-stacked bilayer graphene with radius $R$ for a zero width
subjected to a perpendicular magnetic field $B$. An analytical
approach, using the Dirac equation, is implemented to obtain the
energy spectrum by freezing out the carrier radial motion.
The obtained spectrum exhibits different symmetries and
for a fixed total angular momentum $m$,
it has a hyperbolic dependence of the magnetic field. In
particular, the energy spectra are not invariant under the
transformation $B \longrightarrow  -B$. The application of a
potential, on the upper and lower layer, allows to open a gap in
the energy spectrum and the application of a non zero magnetic
field breaks all symmetries. We also analyze the basics
features of the energy spectrum to show the main similarities and
differences with respect to ideal quantum ring in monolayer,
AB-stacked bilayer graphene and a quantum ring with finite width
in AB-stacked bilayer graphene.




\end{abstract}
\end{center}

\vspace{3cm}

\noindent PACS numbers: 81.05.ue, 81.07.Ta, 73.22.Pr.

\noindent Keywords: Ideal quantum ring, AA-bilayer graphene, magnetic field, energy spectrum.
\end{titlepage}


\section{Introduction}

Graphene, an isolated single layer of graphite, since its
isolation in 2004 \cite{Novoselov04} has attracted many
experimental and theoretical research activities. This lead to
discovery of many interesting properties \cite{Castro09} not been
observed in the ordinary two dimensional electron gas. The large
interest is due to both the unusual mechanical and electronic
properties as well as for the prospects of applications, which may
lead to their use in novel nanoelectronic devices. In addition,
graphene offers the remarkable possibility to probe predictions of
quantum field theory in condensed matter systems, as its
low-energy spectrum is described by the Dirac-Weyl Hamiltonian of
massless fermions \cite{Geim07}. These different properties of
graphene are related to the unusual electronic structure of
graphene, in which the charge carriers behave as massless fermions
with a gapless linear dispersion.

Graphene can not only exist in the free state, but two or more
layers can stack above each other to form what is called few layer
graphene. As example for two coupled graphene sheets, it is known
as bilayer graphene. Bilayer graphene systems show interesting
properties with strong dependence on stacking. In bilayer graphene
there are four atoms per unit cell, with inequivalent sites $A_1$,
$B_1$ and $A_2$, $B_2$ in the first and second graphene layers,
respectively. There are two dominant ways in which the two layers
can be stacked. The first one is the so called AB-stacked bilayer
graphene \cite{Bernal24,Charlier91} and the second is the
AA-stacked bilayer graphene \cite{Lee08,Andres08}. In the
AB-stacking, the layers are arranged in such a way that the $A_1$
sublattice is exactly on top of the sublattice $B_2$. In the
AA-stacking, both sublattices of one sheet $A_1$ and $B_1$, are
located directly on top of the two sublattices of the other sheet
$A_2$ and $B_2$.

Recently, bilayer graphene has surged as another attractive
two-dimensional carbon material and demonstrated new unusual
physical properties
\cite{McCann06,Ohta06,Pereira071,Zhou07,Oostinga08,Zhang09,Kumar11}.
In fact, bilayer graphene is a very different material from
monolayer graphene and also from graphite. The AB-stacked bilayer
graphene has a gapless quadratic dispersion relation, two
conduction bands and two valance bands, each pair is separated by
an interlayer coupling energy of order $\gamma_1 = 400$meV.
However, the energy bands for AA-stacked bilayer graphene are just
the double copies of single layer graphene bands shifted up and
down by the interlayer coupling $\gamma = 200$meV. By applying a
perpendicular electric field on the upper and lower layer, the
spectrum is found to display a gap, which can be tuned by varying
the bias or by chemical doping of the surface \cite{EdMc06}. This
tunable gap can then be exploited for the development of bilayer
graphene devices. In particular, the possibility of controlling
the energy gap has raised the possibility of the creation of
electrostatically defined quantum dots \cite{Pereira07,Pereira09}
and quantum ring \cite{Zarenia09,Zarenia10} in bilayer graphene.

Quantum rings in graphene have also attracted some interest. They
are expected to find application in microelectronics as well as in
future quantum information devices. In fact, a very important
class of quantum devices consists of quantum rings. It have been
studied in semiconductor systems, both experimentally and
theoretically \cite{Fuhrer01}. Recently, quantum rings have been
studied both theoretically and experimentally in monolayer
graphene \cite{Russo08,Huefner10}. The graphene-based quantum
rings have been obtained experimentally by lithographic techniques
\cite{Huefner10}. These systems have been studied theoretically in
monolayer graphene.  Two different ring systems are considered: a
ring with a smooth boundary and a hexagonal ring with zigzag edges
\cite{Recher07}. For AB-stacked bilayer graphene, it was shown
that it is possible to electrostatically confine quantum ring with
a finite width \cite{Zarenia09}.

In this work, we consider a quantum ring in AA-stacked bilayer
graphene in the presence of an external magnetic field. We obtain
analytical expression of the energy spectrum by solving the Dirac
equation and freezing out the carrier radial motion, for zero and
non zero magnetic field. The obtained energy spectrum for ideal
quantum ring will be investigated numerically to underline the
behavior of our system. We investigate the basic features of our
results and compare them with those for ideal ring in monolayer
graphene and AB-stacked bilayer graphene and also for quantum ring
with finite width in AB-stacked bilayer graphene.

The
set of the paper is organized as follows. In section 2, we present
our problem by setting the Hamiltonian describing the system under
consideration. Subsequently, we use
the eigenvalue equation to find the analytic expressions for the
energy spectrum. In section 3, we present our results and give different discussions.
Section 4 provides a summary and conclusions.

\section{Problem setting}

\begin{figure}[h!]
  \centering
  \includegraphics[width=8cm, height=6cm]{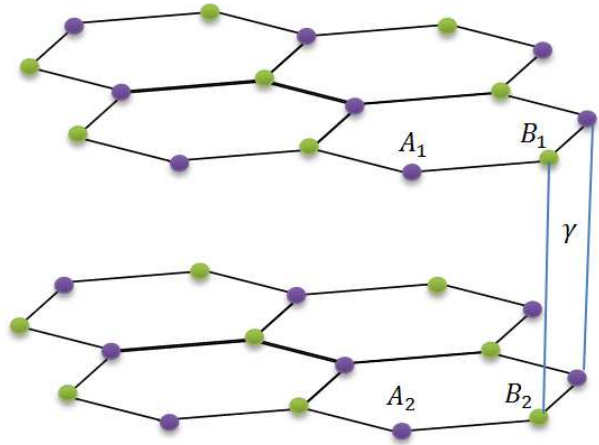}
 \caption{\sf Schematic illustration of lattice structure of AA-stacked
bilayer graphene. It is composed of two graphene layers.}
\label{AA}
\end{figure}

We consider an AA-stacked bilayer graphene quantum ring. This
system is characterized by two monolayer sheets stacked directly
on top of each other. Each carbon atom of the upper layer is
located above the corresponding atom of the lower layer and they
are separated by an interlayer coupling energy $\gamma$ (see
Figure \ref{AA}).

The Hamiltonian in the vicinity of the $K$ and $K'$ valleys, of
the first Brillouin zone, with a perpendicular magnetic field, can
be written as \beq \label{ham1} \mathcal{H}=\begin{pmatrix}
 \tau U & \boldsymbol{\pi}  & \gamma   & 0\\
\boldsymbol{\pi}^{\dagger} & \tau U & 0 &  \gamma\\
\gamma & 0 & -\tau U & \boldsymbol{\pi} \\
0 &  \gamma & \boldsymbol{\pi}^{\dagger} & -\tau U
\end{pmatrix}.
\eeq
Here $\gamma=200$meV is the interlayer coupling term
\cite{Tabert12},
$\boldsymbol{\pi}=v_F(\boldsymbol{p}+e \boldsymbol{A})$
with $\boldsymbol{p}$ being the two-dimensional momentum operator,
in which the symmetric gauge is used to describe the vector
potential $\boldsymbol{A}$, $v_F=10^6$ m/s is the Fermi velocity
 and $\tau=\pm 1$ distinguishes the two $K$
and $K'$ valleys. Moreover, the application of a perpendicular
electric filed creates a potential $+U$ in the upper layer and
$-U$ in the lower layer \cite{Wang11,Wang12,Wang13}.
For the AA-stacked bilayer graphene, the Hamiltonian \eqref{ham1}
acts on a four component spinor
$\left[\Psi_{A_1},\Psi_{B_1},\Psi_{A_2},\Psi_{B_2}\right]^{T}$,
where $\Psi_{A_1 (A_2)}$ and $\Psi_{B_1 (B_2)}$ are the envelope
functions associated with the probability amplitudes of the wave
functions on the $A_1(A_2)$ and $B_1(B_2)$ sublattices of the
upper (lower) layer. Since the total angular momentum operator
$J_z$ commutes with $\mathcal{H}$, then we can construct a common
basis in terms of the eigenspinors, in the polar coordinates such
as $\Psi(r,\theta)=e^{i m \theta} \left[
\Phi_{A_1}(r)e^{i\theta},\ i\Phi_{B_1}(r),\
\Phi_{A_2}(r)e^{i\theta},\ i\Phi_{B_2}(r)\right]^{T}$,
 where $m$ is the angular momentum label. Using the symmetric gauge $\boldsymbol{A}=(0,\frac{Br}{2},0)$, to
write the corresponding momentum operators as
\beqar
\boldsymbol{\pi}&=&v_F e^{i \theta} \left[ -i \hbar \left(
\frac{\partial }{\partial r} + \frac{i \partial }{ r
\partial \theta}\right) + i\frac{e B r}{2}\right] \\ \boldsymbol{\pi}^{\dagger}&=&v_F e^{-i \theta} \left[ -i \hbar
\left( \frac{\partial }{\partial r} - \frac{i \partial }{ r
\partial \theta}\right) - i\frac{e B r}{2}\right]. 
\eeqar

In the next we will present analytical expression of the
eigenstates and energy levels of ideal quantum ring created with
AA-stacked bilayer graphene. For an ideal ring with radius $R$,
the momentum of the charge carriers in the radial direction is
zero. By freezing out the carrier radial motion, the
four-component wave function
becomes \beq \label{Spi2} \Psi(R,\theta)=\begin{pmatrix}
\Phi_{A_1}(R)e^{i\theta} \\
i\Phi_{B_1}(R)\\
\Phi_{A_2}(R)e^{i\theta}\\
i\Phi_{B_2}(R)
\end{pmatrix} e^{im\theta}.
\eeq
By solving the Dirac equation
$\mathcal{H}\Psi(R,\theta)=E\Psi(R,\theta)$, we obtain the
following system of coupled differential equations
\begin{eqnarray}
\label{eq47}
\left\{%
\begin{array}{llll}
(E-\tau U)\Phi_{A_1}(R)-\eta(m+\beta)\Phi_{B_1}(R)+\gamma\Phi_{A_2}(R)&=0  \\
\eta (m+\beta +1)\Phi_{A_1}(R)-(E+\tau U)\Phi_{B_1}(R)-\gamma\Phi_{B_2}(R)&=0\\
\eta (m+\beta +1)\Phi_{A_2}(R)-(E-\tau U)\Phi_{B_2}(R)-\gamma\Phi_{B_1}(R)&=0\\
\gamma \Phi_{A_1}(R)-\eta (m+\beta)\Phi_{B_2}(R)+(E+\tau
U)\Phi_{A_2}(R)&=0
\end{array}%
\right.
\end{eqnarray}
where we have set the quantities \beq \eta=\frac{\hbar v_F}{R},
\qquad \beta =\frac{eB }{2\hbar}R^2.\eeq After some
straightforward algebra, we end up with the  polynomial
equation that determine the energy spectrum 
\beq\lb{eneq}
E^4-2E^2\left(\eta^2 \alpha+U^2 +\gamma^2\right)+\left(\eta^2
\alpha+U^{2}\right)^2 -2\left(\eta^2 \alpha
-U^{2}\right)\gamma^2+\gamma^4=0 
\eeq 
where the parameter $\alpha$ is given by
\beq
\alpha=(m+\beta)^2+(m+\beta). 
\eeq
There are  four
solutions for \eqref{eneq}
\beq \lb{ener} E=s\sqrt{(\gamma \pm \eta\sqrt{\alpha})^{2}+U^2}
\eeq
with $s=\text{sign}(E)$. We notice that for $m+\beta\leq -1$ or
$m+\beta \geq 0$ the four solutions are real, except for
$-1<m+\beta < 0$, which
they are complex.

\section{Results and discussions}

\begin{figure}[h!]
  \centering
  \includegraphics[width=8cm, height=6.1cm]{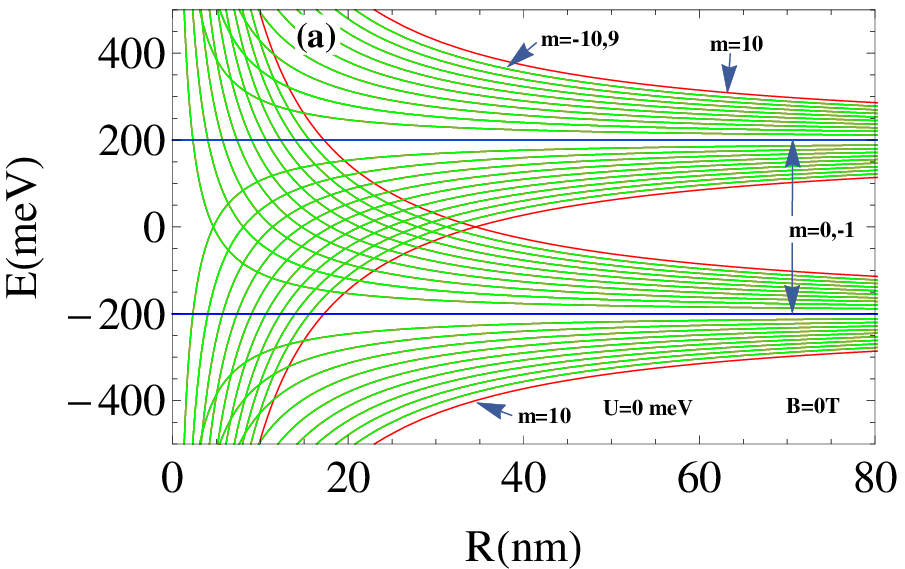}
  \includegraphics[width=8cm, height=6.1cm]{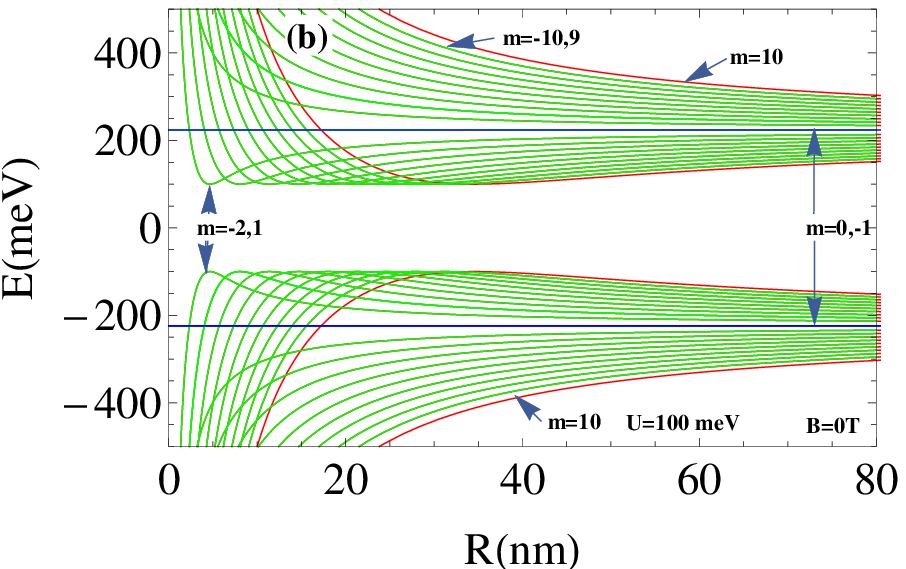}\\
  \includegraphics[width=8cm, height=6.1cm]{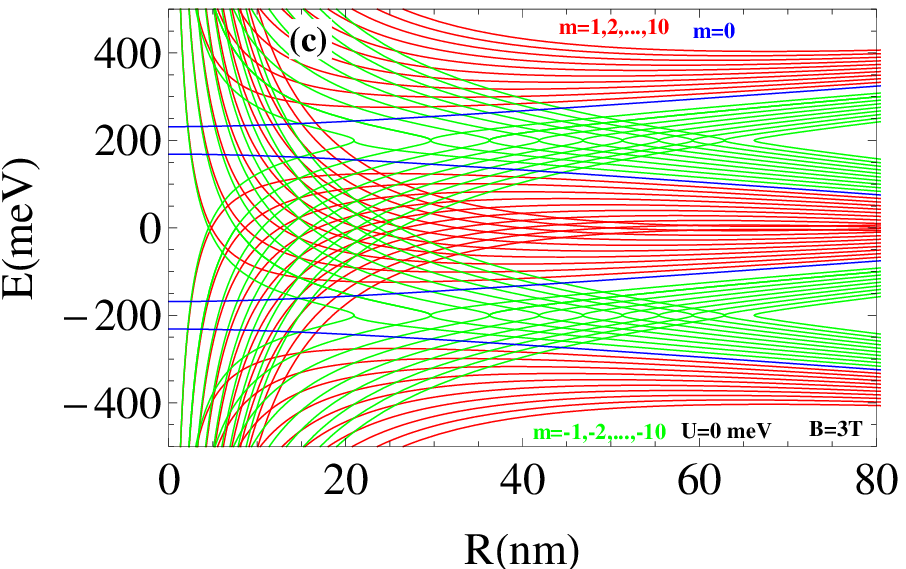}
  \includegraphics[width=8cm, height=6.1cm]{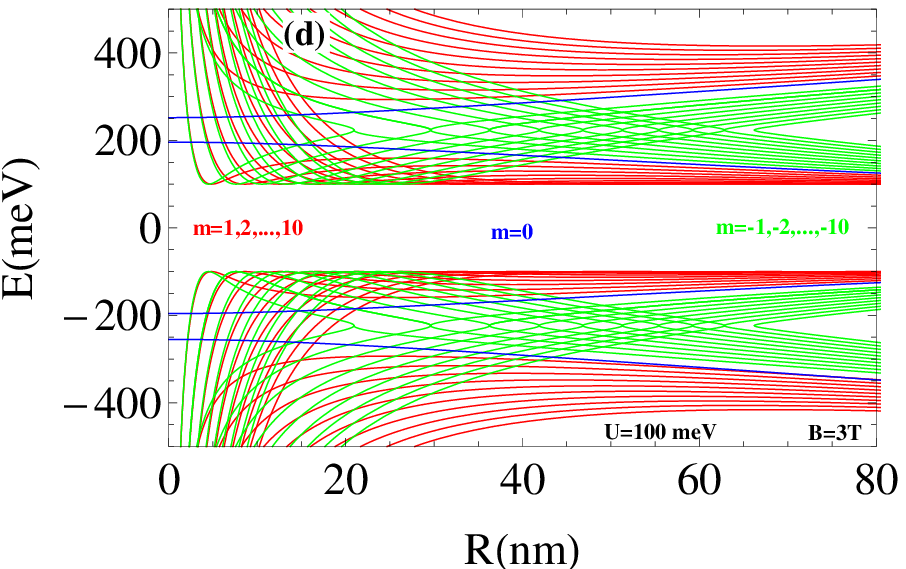}
  \caption{\sf Energy as  function of the radius $R$ for an ideal
quantum ring in AA-stacked bilayer graphene with the angular
quantum number $-10 \leq m \leq -1$ (green curves), $1 \leq m \leq
10 $ (red curves) and $m=0$ (blue curves). (a): Zero magnetic
field $B= 0$T and $U = 0$meV. (b): Zero magnetic field $B = 0$T
and $U = 100$meV. (c):  Non zero magnetic field $B = 3$T and
$U=0$meV. (d): Non zero magnetic field $B = 3$T and $U = 100$meV.}
\label{E-R}
\end{figure}

In Figure \ref{E-R}, we plot the energy levels of an ideal quantum
ring in AA-stacked bilayer graphene as a function of the ring
radius $R$ with $-10 \leq m \leq 10$. The green and red curves
correspond, respectively, to $-10 \leq m \leq -1$ and $1 \leq m \leq
10 $, while the blue one corresponds to $m = 0$. For $U= 0$meV
(Figure \ref{E-R}(a) and (c)), we see that the energy
spectrum shows two set of levels and the energy spectrum for $U=0$meV
resembles those found in the case of AA-stacked bilayer graphene
quantum dot \cite{BZJ16}. It is clear that the two set of this
AA-stacked bilayer are just double copies of the energy spectrum
corresponding to monolayer graphene one, shifted up/down by $= + /- \gamma$. 
We notice that for zero magnetic field, the energy take
the following form 
\beq \lb{eqE}E(m,\beta=0)=s\sqrt{(\gamma
\pm\eta\sqrt{m(m+1)})^{2}+U^2}. 
\eeq

\noindent It is important to note that
for $\gamma = 0$, we
recover the energy of monolayer graphene \cite{Zarenia10}. For
large ring radius $R$, the upper set of levels converges to the
interlayer hopping energy $\gamma= 200$meV. However, the lower set of
levels converges to $\gamma= -200$meV. In addition, we notice that
from
\eqref{eqE} and for $m=0,- 1$ the energy will be
independent of the radius $R$ 
\beq
E=s \sqrt{\gamma^2 + U^2}
\eeq
and therefore
all branches are twofold degenerate. Furthermore,  \eqref{eqE} exhibits 
interesting spectrum symmetries, such as 
\beqar  
E(m,0) &=& E(-m-1,0) \\
E(0,0) &=& E(-1,0). 
\eeqar 
The application of a potential, on the
upper and lower layer, allows to open a gap in the energy spectrum
(Figure \ref{E-R}(b) and (d)). When the ring radius increases, for
zero magnetic field, the gap width increases as well, which can be seen
clearly in Figure \ref{E-R}(b). Note that, in the case of a
quantum ring with finite width \cite{Zarenia09}, the results show
a weak dependence on the ring radius.

Furthermore, the numerical results demonstrate that the
application of a non zero magnetic field ($B=3$T) break the
degeneracy of all branches. In contrast with the results obtained
from the Schr\"{o}dinger equation, the electron and hole energy
levels are not invariant under the transformation
$B\longrightarrow -B$ \cite{Zarenia09}. These results are similar
to those obtained for an ideal quantum ring in monolayer and
AB-stacked bilayer graphene \cite{Zarenia10} and also for quantum
ring with finite width \cite{Zarenia09}. For non zero magnetic
field and for large ring radius $R$,
\eqref{eqE} becomes 
\beq\lb{enma} 
E(m,\beta)=s\sqrt{(\gamma \pm\lambda
R)^{2}+U^2} 
\eeq 
where the parameter $\lambda=\frac{eBv_F}{2}$ is magnetic field dependent. This can clearly
explain the approximately linear dependence of the energy branches
on the ring radius for large $R$. However, for small $R$, \eqref{enma}
reduces to the form
\beq 
E(m,\beta)=s\frac{\hbar v_F}{R}\sqrt{m(m+1)} 
\eeq 
showing that the
spectrum has $1/ R$ dependence. In addition, like the case of
zero magnetic field, the application of a potential open a gap in
the energy spectrum. But, the gap width remains unchanged by
increasing ring radius $R$.

\begin{figure}[h!]
  \centering
  \includegraphics[width=10cm, height=6.5cm]{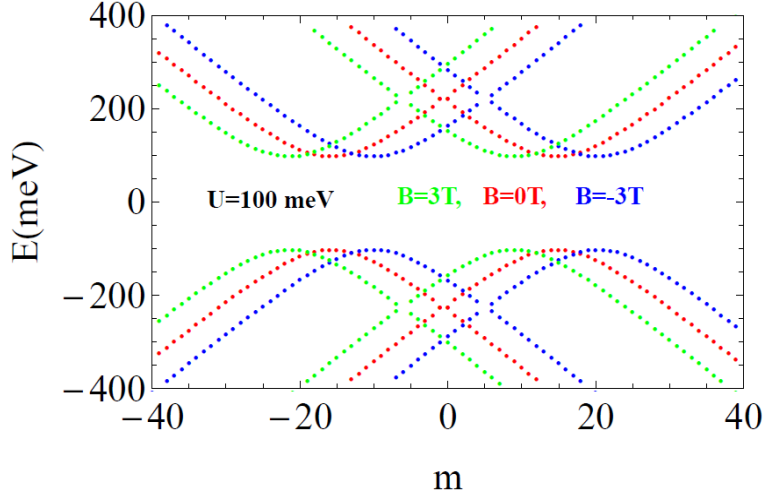}\\
  \caption{\sf Energy levels of an ideal quantum ring in AA-stacked bilayer graphene 
as function of the angular momentum $m$ for $B=$-3T, 0T, 3T,
with $U=100$meV and $R=50$nm.} \label{Em}
\end{figure}

In Figure \ref{Em}, we plot the energy levels of an ideal quantum
ring as function of the angular momentum $m$ for three different
values of the magnetic field ($B=$-3T, 0T, 3T), with
$U=100$meV and $R=50$nm. We notice that, like the case of monolayer
\cite{Zarenia10} and AB-stacked bilayer graphene \cite{Costa14},
the electron energy presents a minimum for a particular value of
the angular momentum $m$. The  minimum energy for $B=-3$T is given
by two values of $m$: $m=20$ and $m=-10$. However, the 
minimum energy for $B=0$T and $B=3$T, are respectively, given by $m= 15,
-16 $ and $m = 9, -21 $. This can be explained by the fact that
from
\eqref{ener}, the spectra are invariant under the transformation
$B\longrightarrow -B$ and $m\longrightarrow -(m+1)$. Thus, the
energies are related by the symmetry relation 
\beq
E(m,B)=E(-m-1,-B)
\eeq 
which is also exists in the case of an ideal
quantum ring in monolayer graphene \cite{Zarenia10}.

\begin{figure}[h!]
  \centering
  \includegraphics[width=8.2cm, height=6.5cm]{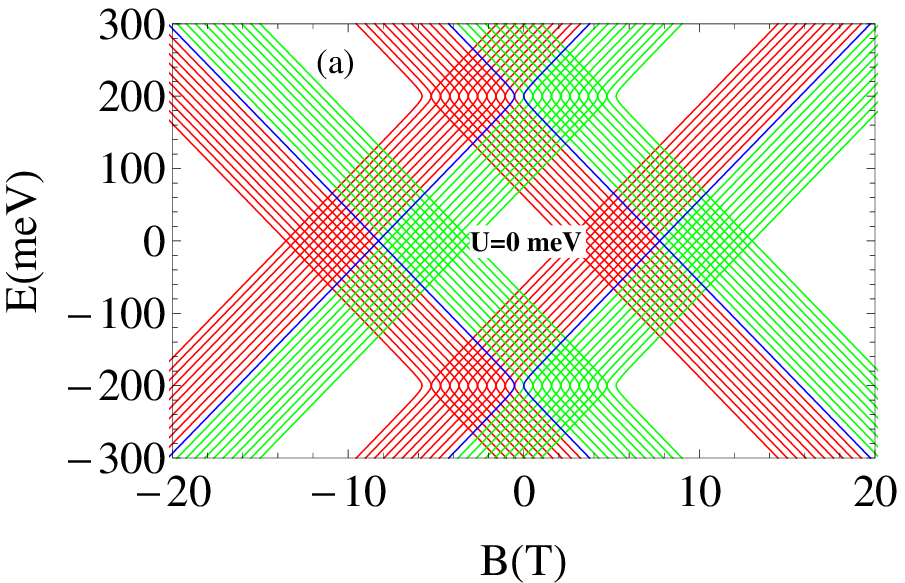}
  \includegraphics[width=8.2cm, height=6.5cm]{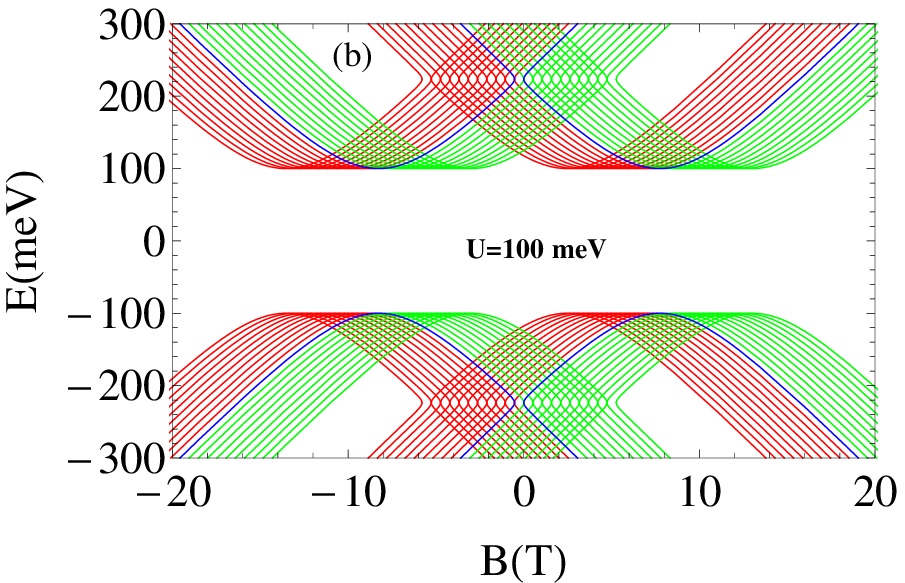}
   \caption{\sf Energy spectrum of an ideal quantum ring in AA-stacked bilayer graphene
    as function of the magnetic field $B$ for $R=50$nm with
    (a): $U=0$meV and (b): $U=100$meV. The energy levels
    are  shown for the quantum number $-10\leq m\leq-1$(green curves),
    $1\leq m \leq 10$ (red curves) and $m=0$ (blue curves).} \label{cc}
\end{figure}

The magnetic field dependence of the energy spectrum is presented  in
Figure \ref{cc} with the ring radius $R= 50$nm and $U=0$meV
(Figure \ref{cc}(a)) and $U=100$meV (Figure \ref{cc}(b)). The
green and red curves, respectively, show the energy for $-10\leq
m\leq-1$ and $1\leq m \leq 10$. However, the blue curves show the
energy for $m=0$. In Figure \ref{cc}(a), we plot the electron and
hole energy levels for ideal quantum ring in AA-stacked bilayer
graphene for $U=0$meV, where
\eqref{ener} reduces to \beq E=s(\gamma \pm
\eta\sqrt{\alpha}).\eeq
We can clearly show that the energy levels are straight lines and
we have zero gap. Moreover, one can see that the energy spectrum
shows two set of levels. They are just the double copies of the
energy spectrum corresponding to ideal quantum ring in monolayer
graphene, one shifted up by $+\gamma$ and other one shifted down
by $-\gamma$, where $ \gamma= 200$meV.
We notice that the energy spectrum resembles those
found in the case of monolayer graphene for ideal quantum ring
\cite{Zarenia10} and for hexagonal ring with zigzag edges
\cite{Recher07}. In Figure \ref{cc}(b), the energy has a
hyperbolic dependence of the magnetic field. In addition, the
application of a potential $U=100$meV leads to the appearance of a
energy gap around the point $E=0$. Also, these results show that the
electron energy spectrum exhibits a minimum at $E=U$ and there is a
symmetry between the electron and hole states. Indeed, the
electron and hole energies are related by the symmetry 
\beq
E_{e}(m,B)=-E_{h}(-m-1,-B) 
\eeq 
where the indices $h$ and $e$
refer, respectively to holes and electrons. These results are not
similar to that obtained for a finite width quantum ring in
AB-stacked bilayer graphene \cite{Zarenia10}, where the electron
energy exhibits two local minima and the electron and hole states
are asymmetric.
\vspace{3mm}

\begin{figure}[h!]
  \centering
  \includegraphics[width=8cm, height=6.7cm]{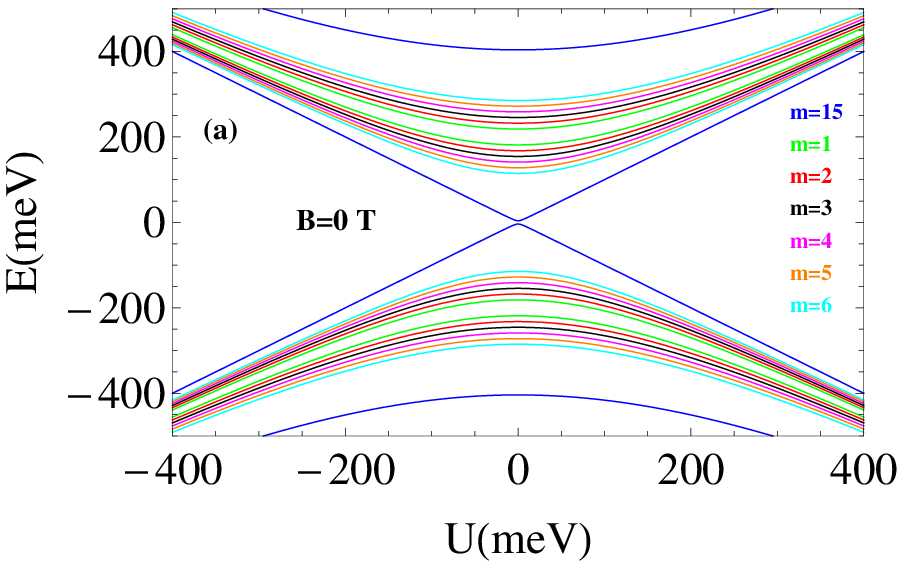}
  \includegraphics[width=8cm, height=6.7cm]{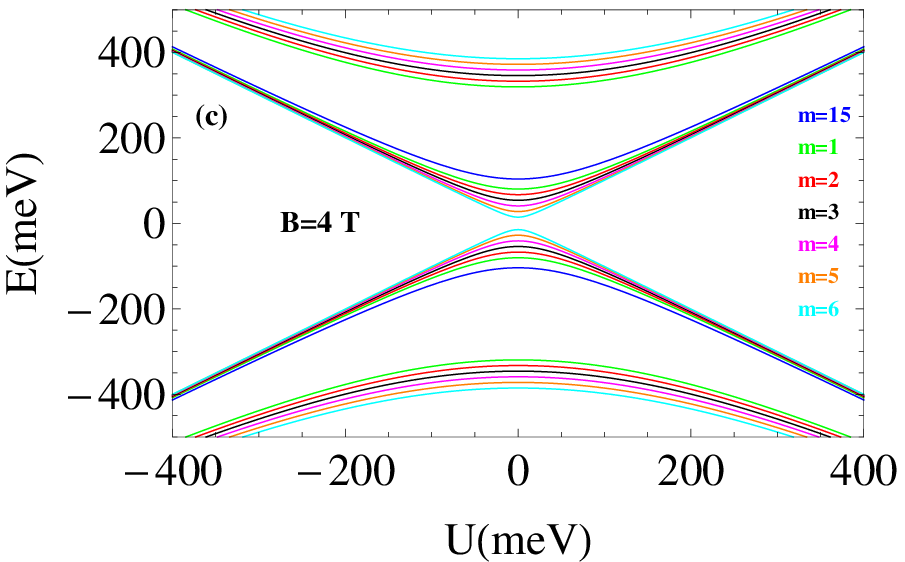}\\
  \includegraphics[width=8cm, height=6.7cm]{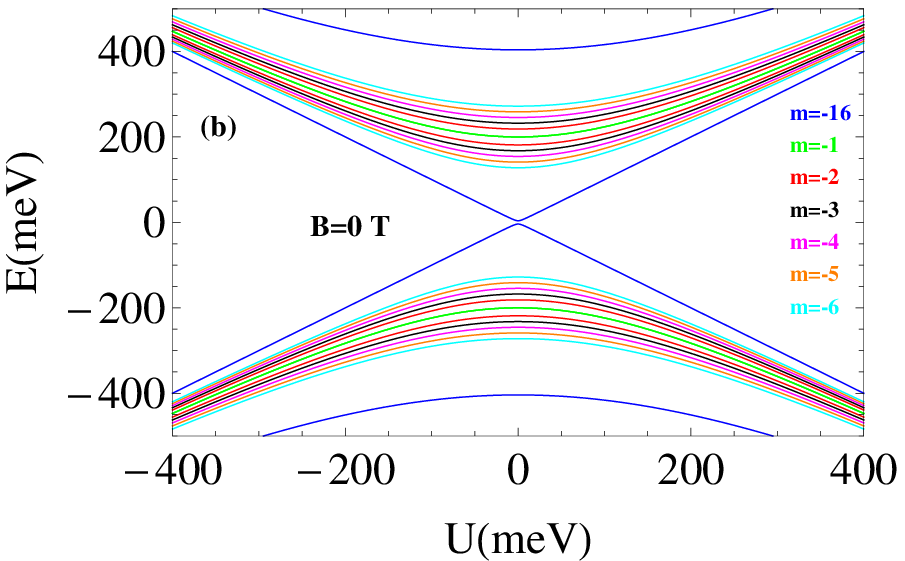}
  \includegraphics[width=8cm, height=6.7cm]{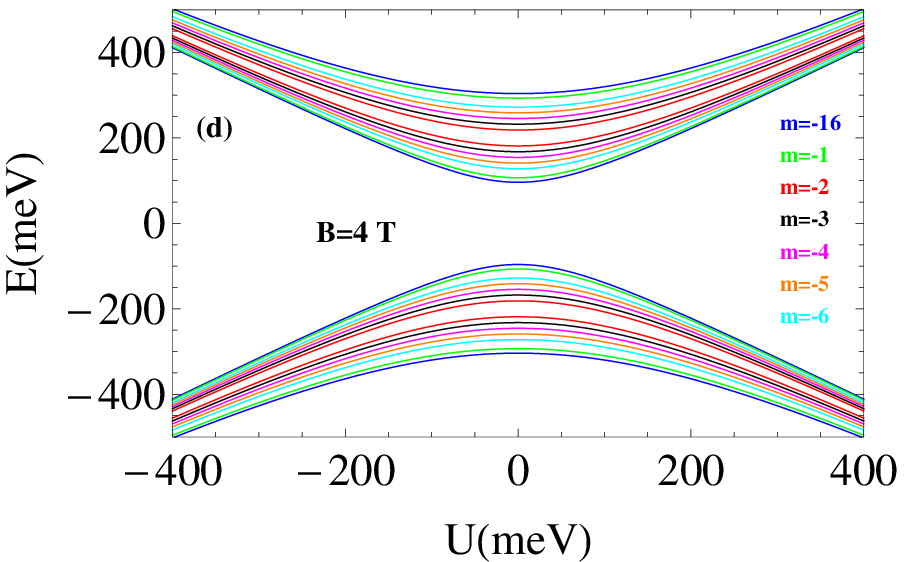}
  \caption{\sf Energy levels of an ideal quantum ring in AA-stacked bilayer graphene
  as function of the potential $U$ with $R=50$nm. (a): $B=0$T and $m > 0$. (b): $B=0$T and $m < 0$. (c): $B=4$T and $m > 0$. (d):
  $B=4$T and $m < 0$.} \label{EU}
\end{figure}

In Figure \ref{EU}, we plot the energy spectrum as a function of
the potential $U$ with $R=50$nm, for zero magnetic field (left
panels) and $B=4$T (right panels) with positive $m$ (upper panels)
and negative $m$ (lower panels). For zero magnetic field the
energy spectrum has an hyperbolic form and is twofold
degenerate due to the fact that we have $E(m)=E(-m-1)$. Note
that an energy gap is opened for non zero magnetic field. These
results are similar to the case of a quantum ring in monolayer
graphene where the energy gap is opened by applying a non zero
magnetic field \cite{Zarenia10}. In the case of AB-stacked bilayer
graphene, the results are similar to those found for a quantum
ring in monolayer graphene where the gate potential has a similar
effect as the mass term. The application of a non zero magnetic
field break the degeneracy.

\section{Conclusion}

In summary, we have investigated a quantum ring in AA-stacked
bilayer graphene by including the effect of an external magnetic
field. The calculation was performed by solving the Dirac equation
for a zero width ring geometry, i.e. ideal ring. In the case of an
ideal ring with radius $R$, the momentum of the carriers in the
radial direction is zero, then we have treat the radial parts of the
spinors as a constant. From the eigenvalue equation, we have
obtained the energy spectrum as function of the ring radius, the
potential and the magnetic field.

Our numerical results showed that the energy spectrum of ideal
quantum ring presents two sets of states as function of the ring
radius $R$. The upper set corresponds to the upper layer and the
lower one corresponds to the lower layer. By increasing the ring
radius, for zero magnetic field, the upper set of levels converges
to the interlayer hopping energy $\gamma=200$meV and the lower set
levels converges to $\gamma=-200$meV. In the absence of the
magnetic field, the energy levels are twofold degenerate where
$E(m,0)=E(-m-1,0)$. It is important to note that the application
of a potential, on the upper and lower layer, allows to open a gap
in the energy spectrum. By increasing the ring radius $R$, the gap
width increases as well. These results differs from previous studies of
graphene-based quantum ring with finite width, where the energy
levels show a weak dependence on the ring radius.

Furthermore, our numerical results demonstrated that the electron
and hole energy levels are not invariant under the transformation
$B\longrightarrow -B$. These results are similar to that obtained
for ideal ring in monolayer and AB-stacked bilayer graphene and
also for a quantum ring with finite width in AB-stacked bilayer
graphene. However, it is not the case with the results obtained
from the Schr\"{o}dinger equation. We notice that the application of
a non zero magnetic field breaks the degeneracy of all branches.
For small ring radius, the energy branches have a $1/R$
dependence. For large ring radius, the branches have an
approximately linear dependence on the ring radius. We have found
also that the field dependence is linear for a fixed total angular
momentum $m$ for $U=0$meV. However, for $U=100$meV, the energy has
a hyperbolic dependence of the magnetic field and exhibits a
minimum for a special values of $m$. In addition, the application
of a potential $U = 100$meV leads to the appearance of a energy gap
around $E = 0$.

\section*{Acknowledgment}

The generous support provided by the Saudi Center for Theoretical
Physics (SCTP) is highly appreciated by all authors.



\begin{thebibliography}{99}

\bibitem{Novoselov04}               K. S. Novoselov, A. K. Geim, S. M. Morozov, D. Jiang, Y. Zhang, S. V. Dubonos, I. V. Grigorieva and A. A. Firosov, Science 306, 666 (2004).
\bibitem{Castro09}                  A. H. Castro Neto, F. Guinea, N. M. R. Peres, K. S. Novoselov and A. K. Geim, Rev. Mod. Phys. 81, 109 (2009).
\bibitem{Geim07}                    A. K. Geim and K. S. Novoselov, Nat. Mater. 6, 183 (2007).

\bibitem{Bernal24}                  J. D. Bernal, Phys. Eng. Sci. 106, 749 (1924).
\bibitem{Charlier91}                J. C. Charlier, J. P. Gonze and X. Michenaud, Phys. Rev. B 43, 6 (1991).
\bibitem{Lee08}                     J.-K. Lee, S.-C. Lee, J.-P. Ahn, S.-C. Kim, J. I. B Wilson and P. John, J. Chem. Phys. 129, 234709 (2008).
\bibitem{Andres08}                  P. L. de Andres , R. Ramirez and J. A. Vergs, Phys. Rev. B 77, 045403 (2008).



\bibitem{McCann06}                  E. McCann, Phys. Rev. B 74, 161403 (2006).
\bibitem{Ohta06}                    T. Ohta, A. Bostwick, T. Seyller, K. Horn and E. Rotenberg, Science 313, 951 (2006).
\bibitem{Pereira071}                J. M. Pereira, Jr., F. M. Peeters and P. Vasilopoulos, Phys. Rev. B 76, 115419 (2007).
\bibitem{Zhou07}                    S. Y. Zhou, G.-H. Gweon, A. V. Fedorov, P. N. First, W. A. de Heer, D.-H. Lee, F. Guinea, A. H. Castro Neto and A. Lanzara, Nature Mater. 6, 770 (2007).
\bibitem{Oostinga08}                J. B. Oostinga, H. B. Heersche, X. Liu, A. F. Morpurgo and L. M. K. Vandersypen, Nature Mater. 7, 151 (2008).
\bibitem{Zhang09}                   Y. Zhang, T. T. Tang, C. Girit, Z. Hao, M. C. Martin, A. Zettl, M. F. Crommie, Y. R. Shen and F. Wang, Nature 459, 820 (2009).
\bibitem{Kumar11}                   S. B. Kumar and J. Guo, Appl. Phys. Lett. 98, 222101 (2011).


\bibitem{EdMc06}                    E. McCann and V. I. Fal'ko,  Phys. Rev. Lett. 96, 086805 (2006).
\bibitem{Pereira07}                 J. M. Pereira, Jr., P. Vasilopoulos and F. M. Peeters,  Nano Lett.  7, 946 (2007).
\bibitem{Pereira09}                 J. M. Pereira, Jr., P. Vasilopoulos, F. M. Peeters and G. A. Farias, Phys. Rev. B 79, 195403 (2009).
\bibitem{Zarenia09}                 M. Zarenia, J. M. Pereira, Jr., F. M. Peeters and G. A. Farias, Nano Lett. 9, 4088 (2009).
\bibitem{Zarenia10}                 M. Zarenia, J. M. Pereira, A. Chaves, F. M. Peeters and G. A. Farias, Phys. Rev. B  81, 045431 (2010).


\bibitem{Fuhrer01}                  A. Fuhrer, S. Lüscher, T. Ihn, T. Heinzel, K. Ensslin, W. Wegscheider and M. Bichier, Nature (London) 413, 822 (2001).
\bibitem{Russo08}                   S. Russo, J. B. Oostinga, D. Wehenkel, H. B. Heersche, S. S. Sobhani, L. M. K. Vandersypen and A. F. Morpurgo, Phys. Rev. B 77, 085413 (2008).
\bibitem{Huefner10}                 M. Huefner, F. Molitor, A. Jacobsen, A. Pioda, C. Stampfer, K. Ensslin and T. Ihn,  New J. Phys. 12, 043054 (2010).
\bibitem{Recher07}                  P. Recher, B. Trauzettel, A. Rycerz, Ya. M. Blanter, C. W. J. Beenakker and A. F. Morpurgo, Phys. Rev. B 76, 235404 (2007).

\bibitem{Tabert12}                  C. J. Tabert and E. J. Nicol, Phys. Rev. B 86, 075439 (2012).
\bibitem{Wang11}                    D. Wang, Phys. Lett. A 375, 4070 (2011).
\bibitem{Wang12}                    D. Wang and G. Jin,  J. Appl. Phys. 112, 053714 (2012)
\bibitem{Wang13}                    D. Wang and G. Jin, Phys. Lett. A 377, 2901 (2013).
\bibitem{BZJ16}                     A. Belouad, Y. Zahidi and A. Jellal, Mater. Res. Express 3, 055005 (2016).

\bibitem{Costa14}                   D. R. da Costa, M. Zarenia, A. Chaves, G. A. Farias and F. M. Peeters, Carbon 78, 392 (2014).





%

\end{thebibliography}
\end{document}